\documentclass[sigconf]{acmart}

\usepackage{balance}

\def\BibTeX{{\rm B\kern-.05em{\sc i\kern-.025em b}\kern-.08emT\kern-.1667em\lower.7ex\hbox{E}\kern-.125emX}}
    
\copyrightyear{2020}
\acmYear{2020}
\setcopyright{acmcopyright}
\acmConference[ASIA CCS '20]{Proceedings of the 15th ACM Asia Conference on
Computer and Communications Security}{October 5--9, 2020}{Taipei, Taiwan}
\acmBooktitle{Proceedings of the 15th ACM Asia Conference on Computer and
Communications Security (ASIA CCS '20), October 5--9, 2020, Taipei, Taiwan}
\acmPrice{15.00}
\acmDOI{10.1145/3320269.3384751}
\acmISBN{978-1-4503-6750-9/20/10}

\DeclareMathAlphabet{\mathcal}{OMS}{cmsy}{m}{n}

\usepackage[T1]{fontenc}
\usepackage{fancyvrb}
\usepackage{graphicx}
\usepackage{amsmath}
\usepackage{amssymb}

\usepackage{mathtools}
\usepackage{xspace}

\usepackage[caption=false]{subfig} 
\usepackage{multirow}
\usepackage{booktabs}
\usepackage{makecell}
\usepackage{url}
\usepackage{clrscode}
\usepackage{latexsym}
\usepackage{siunitx}

\mathchardef\mhyphen="2D          
\usepackage[normalem]{ulem} 
\usepackage{mathpazo}
\usepackage{tgpagella}

\usepackage{crypto-environments}
\usepackage{pifont}

\newcommand{\defeq}{\mathrel{\mathop:}=} 
\newcommand{\getsr}{\gets_{\scriptscriptstyle\$}} 

\newcommand{\cmark}{\ding{51}}%
\newcommand{\xmark}{\ding{55}}%

\newcommand{\size}[1]{\lvert #1 \rvert} 

\newcommand{\prob}[1]{\Pr \left[ #1 \right]}

\newcommand{\adv}{\mathcal{A}} 

\newcommand{\distr}{\mathcal{D}} 
\newcommand{\advdistr}{\distr^\adv} 
\newcommand{\classifier}{\mathcal{C}}
\newcommand{\acc}{\mathrm{acc}} 
\newcommand{\err}{\mathit{err}} 
\newcommand{\error}[1]{\err_{#1}} 
\newcommand{\accuracy}[1]{\acc_{#1}} 
\newcommand{\goodset}[1]{#1_{\text{\cmark}}}
\newcommand{\badset}[1]{#1_{\text{\xmark}}}
\newcommand{\success}{n}
\newcommand{\trials}{N}
\newcommand{\data}{D}
\newcommand{\Xdata}{X^D}

\newcommand{\EVCSA}[1]{\mathrm{EV}\mhyphen\mathrm{CSA}_{#1}} 
\newcommand{\successrate}[1]{\mathbf{succ}^{\mathrm{ev}\mhyphen\mathrm{csa}}_{#1}}
\newcommand{\Advset}{X^{\adv}}

\newcommand{\Classify}{\mathsf{Classify}} 
\newcommand{\Attack}{\mathsf{Attack}} 

\newcommand{\para}[1]{\vspace{1mm} \noindent {\bf #1}}

\newcommand{\ie}{i.e.\@\xspace}
\newcommand{\aka}{a.k.a.\@\xspace}
\newcommand{\eg}{e.g.\@\xspace}

\newcommand{\cf}{cf.\@\xspace}
\newcommand{\vs}{vs.\@\xspace}
\newcommand{\wrt}{w.r.t.\@\xspace}
\newcommand{\etal}{\textit{et al.\@\xspace}}

\newcommand{\our}{Randomized Squeezing}

\newcommand{\featuresq}{Feature Squeezing}
\newcommand{\croprescale}{Cropping-Rescaling}
\newcommand{\regionbased}{Region-Based Classification}

\begin{document}

\fancyhead{}

\title{On the Security of Randomized Defenses\\ Against Adversarial Samples}

\author{Kumar Sharad}
\affiliation{%
  \institution{NEC Labs Europe}
  \city{Heidelberg}
  \country{Germany}
  }
\email{k.sharad@gmail.com}

\author{Giorgia Azzurra Marson}
\affiliation{%
  \institution{NEC Labs Europe}
  \city{Heidelberg}
  \country{Germany}
  }
  \email{giorgia.marson@neclab.eu}

\author{Hien Thi Thu Truong}
\affiliation{%
  \institution{NEC Labs Europe}
  \city{Heidelberg}
  \country{Germany}
  }
\email{hien.truong@neclab.eu}

\author{Ghassan Karame}
\affiliation{%
  \institution{NEC Labs Europe}
  \city{Heidelberg}
  \country{Germany}
  }
\email{ghassan@karame.org}
\begin{abstract}
  Deep Learning has been shown to be particularly vulnerable to adversarial samples. To combat adversarial strategies, numerous defensive techniques have been proposed. Among these, a promising approach is to use randomness in order to make the classification process unpredictable and presumably harder for the adversary to control. In this paper, we study the effectiveness of randomized defenses against adversarial samples. To this end, we categorize existing state-of-the-art adversarial strategies into three attacker models of increasing strength, namely blackbox, graybox, and whitebox (a.k.a.~adaptive) attackers. We also devise a lightweight randomization strategy for image classification based on feature squeezing, that consists of pre-processing the classifier input by embedding randomness within each feature, before applying feature squeezing. We evaluate the proposed defense and compare it to other randomized techniques in the literature via thorough experiments. Our results indeed show that careful integration of randomness can be effective against both graybox and blackbox attacks without significantly degrading the accuracy of the underlying classifier. However, our experimental results offer strong evidence that in the present form such randomization techniques cannot deter a whitebox adversary that has access to all classifier parameters and has full knowledge of the defense. Our work thoroughly and empirically analyzes the impact of randomization techniques against all classes of adversarial strategies.
\end{abstract}



\maketitle

\section{Introduction}

Deep learning (DL) has advanced rapidly in recent years fueled by big data and readily available cheap computation power. Beyond standard machine learning applications, DL has been found extremely useful in numerous security-critical applications such as handwriting recognition, face recognition~\cite{CVPR:TaigmanYRW14}, and malware classification~\cite{SIGCOMM:YuanLWX14,EUROSP:BackesN17,GrossePM0M16}.
When used in such applications, recent studies show that DL is particularly vulnerable to adversarial samples, which are obtained from correctly classified samples by adding carefully selected perturbations to fool classifiers~\cite{Carlini017,PapernotMG16,Goodfellow2015,DBLP:journals/corr/KurakinGB16}. These perturbations are so chosen that they are large enough to affect the model prediction but small enough to go unnoticed (\eg, through a visual check in image recognition applications).
Since DL was shown vulnerable to adversarial samples, numerous attacks and defenses have been developed back and forth~\cite{Xu2017,Papernot2016,MengC17,PapernotM17}.

While these back-and-forth attacks and defenses have clearly advanced state of the art, it is essential to analyze their robustness in different adversarial models to understand how beneficial they are in making DL more robust. Here, we distinguish between three classes of attacker models depending on the adversary's knowledge with regards to the classifier's details:
\emph{blackbox} (\aka~non-adaptive), meaning that the adversary only knows public information, 
\emph{whitebox} (\aka~fully adaptive), i.e., the adversary knows full details of the classifier including any defense in place, 
and \emph{graybox} (\aka~semi-adaptive), reflecting partial knowledge of the classifier's internals.

A popular defensive technique utilizes randomness in the classification process, with the hope to enlarge the search space of successful adversarial perturbations.
The use of randomness to enhance robustness of DL classifiers has been proposed in many different flavors, both at training and classification time, ranging from randomizing the input to modifying the neural network itself in a randomized fashion.
Although many works hint at the potential of such a technique~\cite{ACSAC:CaoG17,ICLR:GuoRCM18,DBLP:journals/corr/abs-1805-04613,xie2018mitigating}, there is still lack of analysis within the community on the robustness of this strategy against state-of-the-art attacks.

\para{Contributions.}
In this work, we study the effectiveness of pre-processing randomized defenses against a wide variety of adversarial strategies, including the strongest whitebox attacks to date. Specifically,
we develop a security model to formally define robustness of machine learning algorithms under the various adversarial strategies that populate the literature (cf.~Section~\ref{sec:model}). Our model, inspired by cryptographic definitions of security, is generic and captures a broad variety of machine learning classifiers.
To investigate the effectiveness of randomization on the classifier's robustness, we present a lightweight defensive strategy, \emph{\our}, that combines the prominent pre-processing defense Feature Squeezing~\cite{Xu2017} with input randomization (cf.~Section~\ref{sec:solution}). We also compare \our{} with two other instantiations of randomized pre-processing techniques: the \emph{\croprescale{}} defense of Guo~\etal~\cite{ICLR:GuoRCM18}, and \emph{\regionbased{}} by~Cao and Gong~\cite{ACSAC:CaoG17}.
We empirically compare the effectiveness of \our{}, \croprescale{}, and \regionbased{}, against state of the art attack strategies to generate adversarial samples from MNIST, CIFAR-10, and ImageNet datasets (cf.~Section~\ref{sec:eval and results}).

Our proposal embeds randomness within the input to the classifier, operating on every pixel of the image independently by adding randomly chosen perturbations to all pixels, prior to applying the squeezing function. The combination of input randomization and squeezing instantiates a specific pre-processing transformation, similarly to~\croprescale{} and~\regionbased.
For all three randomized techniques, our empirical evaluation shows that introducing an appropriate amount of randomness at pixel level does not significantly hamper accuracy and, in case of~\our, it also improves robustness against graybox adversaries~\cite{Goodfellow2015,Carlini017,DBLP:journals/corr/KurakinGB16,DBLP:conf/cvpr/Moosavi-Dezfooli16,DBLP:conf/eurosp/PapernotMJFCS16,WOOT:HeWCCS17} compared to deterministic \featuresq{}.
Our results further highlight that, despite the perturbation induced by randomizing test images, \regionbased{} and~\our{} can achieve high accuracy and robustness without transforming training samples.
This is in contrast to prior findings~\cite{ICLR:GuoRCM18} hinting that input transformation can be effective against adversarial samples, provided that the same transformation is also applied at training time, and opens the possibility to leverage randomness to realize \emph{online} defenses---which can enhance the robustness of pre-trained classifiers in a flexible and efficient manner.
  
To further evaluate the three defenses in the whitebox model, we consider the strongest currently known attacks, tuned for each defense: the Backward Pass Differentiable Approximation (BPDA)~\cite{ICML:AthalyeC018} and the Expectation Over Transformation (EOT)~\cite{ICML:AthalyeEIK18}.
Our results indicate that while these adaptive attacks defeat all three defenses, increasing the amount of randomness used by the defense results in a higher number of iterations, respectively, larger perturbations, necessary for the attacks to succeed.
This  suggests that even in the case of whitebox attacks, randomness may have a positive, although limited, impact on the classifier's robustness---in the sense of forcing the BPDA and EOT attacks to invest greater effort to craft high-confidence adversarial samples.
Our results also support the intuition that introducing unpredictability to the classification process makes it more challenging, for state-of-the-art adaptive attacks, to find adversarial perturbations which are successful regardless of the randomness.

It is therefore plausible that some DL applications, which can reasonably constrain the attacker by limiting the distortion and/or requiring that adversarial samples be generated in real time, may safely employ randomized classifiers even against (properly constrained) whitebox attacks.

As far as we are aware, this is the first work that comprehensively analyses the impact of randomness on DL classifiers under all state-of the art adversarial strategies, covering also the whitebox attacker model.

\section{Background}\label{sec:background}

In this section, we introduce notation and relevant concepts for the subsequent sections.

\para{Notation \& Conventions.}
Let~$X$ be a (finite) set, and let~$\distr\colon X \to [0,1]$ be a probability distribution.
We denote by $x \gets_\distr X$ the random sampling of an element~$x$ according to distribution~$\distr$; we write~$x \getsr X$ for sampling~$x$ uniformly at random.
We denote by~$f\colon X \to Y$ the function defining the classification problem of interest (\aka~``ground truth''), where~$X$ and~$Y$ are the sets of instances and of corresponding classes (or labels), respectively.
A machine-learning classifier~$\classifier$ is an algorithm that aims at emulating function~$f$. 
Typically, the classifier is deterministic and can be thus thought of as a function itself.
In this work, we cover a broader class of classifiers and let~$\classifier$ be any, possibly randomized algorithm.
If $\classifier$ is randomized, we write $y \getsr \classifier(x)$ to denote that on input~$x$ the classifier, run on freshly sampled randomness, outputs label~$y$.
Let $\classifier\colon X \to Y$ be a deterministic classifier.
For~$X'\subseteq X$ we denote by~$\goodset{X'}(\classifier) = \{ x\in X'  :  \classifier(x) = f(x) \}$ the set of instances in~$X'$ where~$\classifier$ agrees with~$f$.
Similarly, we denote by~$\badset{X'}(\classifier) = \{ x\in X'  :  \classifier(x) \neq f(x) \}$ the set of misclassified instances.
Using this notation, we measure a classifier's \emph{(empirical) accuracy}, respectively, \emph{(empirical) error} \wrt a given dataset~$\data = \{ (x,f(x)) : x \in\Xdata\}$, for some~$\Xdata \subset X$ as 
$\accuracy{\Xdata}(\classifier) \defeq \size{\goodset{\Xdata}(\classifier)}/\size{\Xdata}$
and 
$\error{\Xdata}(\classifier) \defeq \size{\badset{\Xdata}(\classifier)}/\size{\Xdata}$, respectively.
For randomized classifiers, the definitions of accuracy and error need to be augmented for incorporating the randomness of the classifier.
Note that accuracy and error can also be used to capture the performance of a classifier \wrt an \emph{adversarially chosen} distribution~$\advdistr$, respectively, input set~$X^\adv$. 

\subsection{Adversarial Samples}\label{sec:background:attacks}

The accuracy of a classifier is measured \wrt samples drawn from a `natural' distribution~$\distr\colon X \to [0,1]$ over the input space.
This approach is grounded in results from computational learning theory~\cite{CACM:Valiant84},
which guarantee a low classification error as long as samples used at test time originate from the \emph{same distribution} of the training samples.
While this assumption may be realistic in a pure machine-learning setting, it is hard to justify in general.
In cybersecurity, e.g., the ``test samples'' are generated by an adversary~$\adv$ attempting to bypass an ML protected system, and may thus be specifically crafted to deviate from the training samples.
This state of affair has been confirmed by the recent advances in attacking ML systems through \emph{adversarial samples}~\cite{Szegedy2014}.

An adversarial sample~$x'$ is derived from a labeled sample~$(x,y)$ by slightly perturbing~$x$, so that~$x'$ still belongs to the original class~$y$, yet it is classified wrongly.
Formally: $x$ and $x'$ have a relatively small distance~$d(x,x') \leq \epsilon$, $f(x') = y$, and $\classifier(x') \neq y$. The tolerated amount of perturbation $\epsilon$ is called \emph{distortion} (\aka~adversarial budget).
The three most common metrics to measure the distance between an adversarial sample~$x'$ and its legitimate counterpart~$x$ are based on the~$L^{p}$-norms ($L^{0}$, $L^{2}$, and $L^{\infty}$): (i) $d^{0}(x,x') = \lvert \{ i : x_i - x'_i \neq 0\}\rvert$, based on the number of modified pixels; (ii) $d^{2}(x,x') = \bigl(\sum_i (x_i - x'_i)^2\bigr)^{\frac{1}{2}}$, based on the Euclidean distance; (iii) $d^{\infty}(x,x') = \max_i (x_i - x'_i)$, based on the maximum difference between pixels at corresponding positions, where $x_i - x'_i$ is the difference between pixels at position~$i$ of images~$x$ and~$x'$, respectively.
For a distance metric~$d^{p}$, we denote by~$|| \cdot ||_p$ the corresponding norm.
Depending on the attacker's goal, adversarial samples can be categorized as \emph{targeted} and \emph{untargeted}.
A targeted adversarial sample~$x'$ is successful if the classifier's prediction matches an attacker-chosen label~$y_t \neq y$, where~$y$ is the true label.
An untargeted adversarial sample instead succeeds if the classifier predicts any label other than~$y$.

Prominent techniques to generate adversarial examples against DL classifiers are the Fast Gradient Sign Method (FGSM)~\cite{Goodfellow2015}, the Basic Iterative Method (BIM)~\cite{DBLP:journals/corr/KurakinGB16}, the Jacobian Saliency Map Approach (JSMA)~\cite{DBLP:journals/corr/KurakinGB16}, the Carlini-Wagner (CW) attacks~\cite{Carlini017}, and DeepFool~\cite{DBLP:conf/cvpr/Moosavi-Dezfooli16}.
Among others, we will consider these attack strategies in our evaluation (\cf~Section~\ref{sec:eval and results}).

\subsection{Defensive Techniques}\label{sec:background:defenses}

Here we discuss the defenses against adversarial samples which are most relevant to our work. We survey more defensive techniques in Section~\ref{sec:relatedwork}.

\para{Feature Squeezing.}
This technique, introduced by Xu~\etal~\cite{Xu2017}, transforms the input by reducing unnecessary features while keeping the DL model intact. Feature Squeezing is a generic transformation technique to reduce feature input space such that it can limit opportunities for an adversary to generate adversarial samples. The approach assumes that legitimate samples have same output on original and transformed form while adversarial samples have larger difference on outputs, the discrepancy of outputs helps to reject adversarial samples. In this paper, we study the two proposed squeezing techniques, \textit{squeezing color bit depths} and \textit{spatial smoothing}.

Squeezing color bits relies on the assumption that large color bit depth is not necessary for a classifier to interpret an image. The authors consider 8-bit gray scale images of size $28 \times 28$ pixels (MNIST dataset) and 24-bit color images of size $32 \times 32$ pixels (CIFAR-10 and ImageNet datasets) in their experiments. The 8-bit gray scale images are squeezed to 1-bit monochrome images by using a binary filter with cut-off set to 0.5, while each channel of the 24-bit color images (8-bit per color channel) is squeezed to 4 or 5 bits. Each channel can be reduced to $i$-bit depth by multiplying the input value with $2^{i}-1$, rounded up to integers and then divided by $2^{i}-1$ to scale back to [0,1].

The local smoothing is a type of spatial smoothing technique that adjusts the value of each pixel based on aggregated values, \eg, by taking median of its neighborhood pixels. The median smoothing technique follows Gaussian smoothing design. The values of neighborhood pixels are decided by a configurable sliding window of which size ranges from 1 to entire image. Experiments in~\cite{Xu2017} show that median smoothing with $2 \times 2$ and $3 \times 3$ sliding window is effective. Another way to perform spatial smoothing is non-local smoothing. Non-local smoothing considers a large area to compute replacement value for each pixel. Given an image patch, non-local smoothing searches for all similar patches and replaces the center patch with the average of similar patches. We use the notation proposed by Xu~\etal~\cite{Xu2017} to denote a filter as ``non-local means (a-b-c)'', where $a$ is the search window $a \times a$, $b$ the patch size $b \times b$ and $c$ the filter strength.

\para{Randomness-Based Defenses.} The literature features a number of strategies to use randomness for enhancing DL classifiers against adversarial samples.
Zhou~\etal~\cite{DBLP:journals/corr/abs-1805-04613} propose two ways to use randomness for strengthening deep neural-network (DNN) models: to add random noise to the weights of a trained DNN~model, and to select a model at random from a pool of train DNN~models for each test input.
Xie~\etal~\cite{xie2018mitigating} use randomness in a different way:
to resize the image to a random size, or to add padding zeroes in a randomized fashion.
We discuss two other existing randomization strategies which will be later considered in our evaluation, namely \emph{region-based classification}~\cite{ACSAC:CaoG17} and \emph{cropping-rescaling}~\cite{ICLR:GuoRCM18}, along with our proposal in Section~\ref{sec:solution}.

\section{Security Model}\label{sec:model}

In this section, we present a security model for evasion attacks that allows us to formalize \emph{robustness} to adversarial samples.

\para{Game-based Modeling of Evasion Attacks.} Our model considers an adversary~$\adv$ that aims at defeating a classifier~$\classifier$ by generating adversarial samples starting from ``natural'' samples.
Following the approach of modern cryptography, our security model reproduces the above scenario through a security game between~$\adv$ and~$\classifier$, that we name \emph{evasion under chosen-sample attacks} (EV-CSA), as illustrated in Figure~\ref{fig:securitygame:evasion}.

\begin{figure}[ht]
  \begin{center}
    \resizebox{0.9\linewidth}{!}{%
      \fbox{
        \begin{minipage}[t]{6cm}
        \begin{experiment}{$\EVCSA{\adv,\epsilon,\trials}(\classifier,X^\data)$}
        \item $q \gets 0,\ \success \gets 0$
        \item $\Advset \gets \emptyset$
        \item $\adv(\epsilon,N,\langle \mathcal{C}\rangle,X^\data)^{\Classify,\Attack}$
        \item Return $\success/\trials$
          \label{line:endgame}
        \end{experiment}
      \vspace{2mm}	
    \begin{oracle}{$\Classify(x)$}
        \item $\hat y \gets \classifier(x)$
        \item Give $\hat y$ to $\adv$
        \end{oracle}
        \vspace{2mm}	
        \begin{oracle}{$\Attack(x,x',y_t)$}
        \item If $q \geq \trials$: Go to line~\ref{line:endgame}
        \item Enforce  $(x,*) \notin \Advset$
          \label{line:advsample:freshness}
        \item $q \gets q +1$
        \item $\Advset \stackrel{\cup}{\gets} (x,x')$ 
        \item If $\classifier(x') = y_t$ and $d(x',x) \leq \epsilon$:%
          \label{line:adv:sample:condition}
        \item \quad $\success \gets \success +1$%
          \label{line:successful}
        \item Return
        \end{oracle}
    \end{minipage}
     }
  }
 \end{center}
  \caption{
    Security game for targeted evasion under chosen-sample attacks (EV-CSA), involving adversary~$\adv$ against classifier~$\classifier$.
    Untargeted attacks are captured by replacing the inputs to the~$\Attack$ oracle with pairs~$(x,x')$ and, the first condition of line~\ref{line:adv:sample:condition} with $\classifier(x') \neq \classifier(x)$. 
  }
  \label{fig:securitygame:evasion}
\end{figure}

The adversary's goal is to present a number of adversarial samples generated from a set~$X^\data \subset X$ of ``naturally occurring'' (labeled) samples.
\footnote{In practice, $X^\data$ represents a set of available images used for testing, e.g., MNIST.}
The number~$N$ of adversarial samples, $1 \leq N \leq \size{X^\data}$, is a game parameter and can be adapted to capture different security goals.

We specify the amount of information that~$\adv$ has about the adversarial task by passing the relevant inputs:
the allowed adversarial perturbation (\aka~distortion)~$\epsilon$,
the number~$N$ of adversarial samples,
the classifier's code~$\langle \classifier \rangle$,
and the set~$X^\data$ of benign samples.
Further, by limiting the amount of information encoded in~$\langle \classifier \rangle$, our game can cover different adversarial models such as ``whitebox'' (\aka~fully adaptive), meaning that~$\adv$ knows every detail about the classifier, including neural-network weights and any defense mechanism in place, ``blackbox'' (\aka~non adaptive), i.e., $\adv$~knows only public information about~$\classifier$,
and intermediate attacker's models, so-called ``graybox'' (\aka~semi-adaptive), in which $\adv$ has only partial information about the classifier's internals and/or defensive layers.
Graybox attacks include those agnostic of a defense mechanism. In this case, the adversary knows the original classifier fully, hence it is not blackbox, but it does not know the defense, hence it is not whitebox either (\cf~Section~\ref{sec:attack:categorization}).

We further let the adversary interact with the classifier through an oracle~$\Classify$,
i.e., $\adv$ can query~$\classifier$ on any input~$x$ of their choosing and obtain the corresponding label~$\hat y = \classifier(x)$.
Observe that the~$\Classify$ oracle provides no extra power to whitebox adversaries, as having full knowledge of the classifier allows~$\adv$ to emulate the oracle.
It is, however, necessary to cover weaker attacks, such as transferability attacks~\cite{PapernotMG16} (which are blackbox), and attacks oblivious of the defense (which are graybox).

The game also provides the adversary with a second oracle, denoted by~$\Attack$, which lets~$\adv$ submit candidate adversarial samples. This oracle allows us to describe $\adv$'s goal formally and to define \emph{robustness} to adversarial samples, as we see next.
The adversary can present an adversarial sample by submitting a query~$(x,x',y_t)$ to the~$\Attack$ oracle, where~$x$ is the starting sample, $x'$ is the candidate adversarial sample, and~$y_t$ is the target label.
Upon being queried, the oracle then checks whether the adversary reached the query limit~$q \geq N$, terminating the game in such a case (\cf~line~\ref{line:endgame}).
Otherwise, it checks whether the adversarial sample~$x'$ is ``fresh'', in the sense that no other adversarial sample~$x''$ has been already proposed for the same starting image~$x$ (\cf~line~\ref{line:advsample:freshness}), which is necessary to invalidate trivial attacks that artificially achieve high success rate, e.g., by presenting ``the same'' adversarial sample over and over, in a trivially modified version.%
\footnote{Changing a few pixels of a successful adversarial sample~$x'$ yields a new sample~$x'' \neq x'$ which is very likely to also be successful, thus $\adv$ should only get credit for one of them.}
If the query gets through the checks, the oracle adds the fresh pair~$(x,x')$ to the adversarial set~$X^\adv$, and further checks whether the classifier errs on~$x'$ as desired, by predicting its class as~$y^t$, and whether~$x'$ is sufficiently close to~$x$, i.e., $d(x,x') \leq \epsilon$ according to some pre-established distance metric~$d$.
In case of success, the game rewards the adversary by increasing the counter~$n$ which records the number of successful samples (\cf~line~\ref{line:successful}).

As soon as the~$\trials$ ``chosen samples'' are submitted, the EV-CSA game ends outputting the success rate of the adversary, that we denote by~$\successrate{\adv,\epsilon,\trials}(\classifier,X^\data) = \success/\trials$.
An execution of the EV-CSA game depends on the adversarial strategy~$\adv$ and the classifier~$\classifier$---both of which may be randomized.
In particular, if the game depends on any randomness (used by the adversary, by the classifier, or both), the outcome is determined by the value of the randomness, and the success rate is a random variable.

\para{Deterministic \vs Randomized Classifiers.}
We stress that oracle~$\Attack$ does not reflect an actual capability of the attacker, however, it provides a natural abstraction for determining $\adv$'s success rate.
In particular, if only deterministic classifiers were considered, having~$\adv$ submit their samples through the oracle is equivalent to letting~$\adv$ present a set~$\Advset$ of~$\trials$ samples directly.
That is, the usual notion of success rate against deterministic classifiers is a special case of our notion.
The reason for introducing the~$\Attack$ oracle is precisely that, when \emph{randomized} classifiers are considered, it is no longer meaningful to talk about a \emph{set} of adversarial examples
(a given sample~$x'$ may be correctly labeled for some choices of $\classifier$'s randomness while being misclassified for a different randomness).

\para{Defining Robustness.}
Our security game provides a formal language to express the effectiveness of a defense in making a given classifier ``more robust'' to attacks.
For a classifier~$\classifier$, let~$\classifier^d$ denote the classifier obtained from~$\classifier$ by applying a defense~$d$.
Intuitively, a defense is effective against an attack~$\adv$ if either $\adv$'s success rate after applying the defense is significantly smaller than that with no defense, or a larger distortion is necessary to achieve that success rate.
Formally, we say that a defense~$d$ for classifier~$\classifier$ is \emph{effective} against attack~$\adv$,
or equivalently that $\classifier^d$ is \emph{more robust} than~$\classifier$, 
if either $\successrate{\adv,\epsilon,\trials}(\classifier^d) \ll \successrate{\adv,\epsilon,\trials}(\classifier)$, or $\successrate{\adv,\epsilon^d,\trials}(\classifier^d) = \successrate{\adv,\epsilon,\trials}(\classifier)$ for $\epsilon^d \gg \epsilon$.

\section{Randomness-Based Defenses}\label{sec:solution}

In this section, we present the three randomness-based defensive techniques which we consider in our empirical evaluation from Section~\ref{sec:eval and results}.
While they all apply a randomized pre-processing layer at test time,
the first defense, \croprescale{}~\cite{ICLR:GuoRCM18}, also operates on the training phase, thereby leading to an ``offline'' defense.
It further uses an ensembling technique, meaning that classification is based on the model predictions over an ensemble of samples generated from the original input.
The second defense, \regionbased{}~\cite{ACSAC:CaoG17}, 
does not modify the training phase---i.e., it is ``online''---but uses ensembling, too.
The third defense, \our{} (our design), neither alters training nor relies on ensembling. Instead, it combines the randomness layer with a subsequent image-denoising operator based on \featuresq{}~\cite{Xu2017}.

\subsection{Cropping-Rescaling}\label{sec:sol:cropping-rescaling}

The \emph{\croprescale{}} defense by Guo~\etal~\cite{ICLR:GuoRCM18} applies a randomized transformation that crops and rescales the image prior to feeding it to the classifier. The intuition behind the defense is to alter the spatial positioning of the adversarial perturbation, so that it no longer causes the desired effect and, therefore, it makes the corresponding adversarial sample less likely to succeed.
More precisely, cropping-rescaling operates in two steps, at training and at test time.
Training is performed on cropped and rescaled images, following the data augmentation paradigm of He~\etal~\cite{DBLP:conf/cvpr/HeZRS16}.
Then, to predict the label of each test image, the classifier randomly samples 30 crops of the input, rescales them, and averages the model predictions over all crops. 
Applying the input transformation also at training yields higher classification accuracy on adversarial samples~\cite{ICLR:GuoRCM18}.

\subsection{Region-Based Classification}\label{sec:sol:regionbased}

The \emph{\regionbased{}} defense proposed by Cao and Gong~\cite{ACSAC:CaoG17} computes each prediction over an ensemble generated from the input sample in a randomized fashion.
Specifically, this approach samples~10,000 images uniformly at random from an appropriately sized hypercube centered at the testing image, invokes a DNN to compute predictions over the sampled images, and returns the label predicted for the majority of the images---therefore, classification is no longer ``point-based'' but ``region-based''.
Here, an ``appropriate size'' of the hypercube is chosen so that the region-based classifier maintains the accuracy of the underlying DNN over a (benign) test set.
Taking the ``majority vote'' over the ensemble predictions is based on the assumption that while for benign images most neighboring samples yield the same predicted label, adversarial samples are close to the DNN's classification boundary. 

\subsection{\our{}}\label{sec:solution:our}
 
The defensive strategy that we propose, \emph{\our}, combines input randomization with a deterministic image-denoising technique, namely the~\featuresq{} defense by Xu~\etal~\cite{Xu2017} (\cf~Section~\ref{sec:background:defenses}).
We introduce randomness at feature level, for each feature component and within a predefined threshold, so that it does not bias the prediction excessively in any particular direction.
Concretely, let~$\classifier$ denote a DL classifier enhanced with \featuresq{}.
Our proposal preprocesses~$\classifier$'s input by adding a perturbation~$rand$, chosen uniformly at random from the real interval $[-\delta, +\delta]$, $\delta \in [0,1]$, to each pixel.
The intuition here is that adding a small, carefully crafted perturbation preserves the classifier's output on genuine images, and it significantly affects predictions on adversarial images.
While the randomness added at individual feature level does not destroy the patterns of the pixels, which is critical for correct classification, it does introduce a source of unpredictability in the defense mechanism which enlarges the search space of the adversary. Indeed, to craft a successful adversarial sample, the adversary now has to search for a perturbation that yields the desired prediction for (most of) the various possible randomness values, which is a considerably harder task than fooling (deterministic)~\featuresq.
The increased robustness achieved by randomized classifiers clearly depends on the quality of the randomness, which should be unpredictable from the adversary's perspective.
Thus, it is crucial for security that the random noise be generated from a high-entropy key to seed the underlying cryptographic pseudo-random generator.
More specifically, \our{} comprises of the following subroutines:

\para{Setup:} This procedure performs any instruction needed to initialize the original system.
In addition, it sets the value~$\delta \in [0,1]$ for the magnitude of the randomness
(setting $\delta = 0$ leaves the input unchanged, while $\delta = 1$ is almost equivalent to generating a fresh input uniformly at random),
and initializes the random number generator.
The perturbation magnitude~$\delta$ should be sufficiently large to be effective against adversarial samples, and at the same time be sufficiently small to preserve the classifier's accuracy on normal samples.
In Section~\ref{sec:eval and results}, we analyze in details how to choose~$\delta$ in order to establish a good tradeoff between the achieved accuracy and robustness.

\para{Training:} Since our defense mechanism does not affect the training phase, this step is the same as for the original system.
Upon completion of this phase, we can assume a trained (deterministic) classifier~$\classifier$, based on~\featuresq, which we will use as a basis for our randomized classifier~$\classifier_\$$, as we see next.
    
\para{Classification:}
Upon receiving an input image~$x$, the randomized classifier~$\classifier_\$$ selects a uniformly random key~$k_s$ to seed the underlying pseudo-random generator and expands~$k_s$ until a sufficient amount of (pseudo)randomness has been generated to randomize all pixels of~$x$.
The randomization of each pixel consists in adding a random value~$rand\in[-\delta,+\delta]$.
When the value of the pixel goes outside the allowed intensity threshold (normalized to [0,1] in our experiments), we clip them at the edges instead of taking a modulo and wrapping around.
This is performed to bias the randomness for pixels that are close to the intensity thresholds, which helps to preserve accuracy of the classifier for legitimate samples. The process is repeated for every color channel. Hence, for grayscale images, we add randomization just once as there is only one channel, while for color RGB images randomness is added three times, once for each channel (i.e., ``R'', ``G'', and ``B'', respectively), individually per pixel.
The pre-processing routine of \our{} is depicted in Figure~\ref{fig: pixel randomization}. 
Finally, the resulting image~$x'$ is fed to the (deterministic) classifier~$\classifier$, and the resulting prediction~$\hat y = \classifier(x')$ is returned as label for~$x$.


   \begin{figure}[!htb]
     \centering
     \resizebox{\linewidth}{!}{%
       \fbox{
         \begin{minipage}{\linewidth}
           \begin{codebox}
             \Procname{$\proc{$Randomize_{pixels}$}(Pixels, \delta)$}
             \zi $k_s \gets_\$ \{0,1\}^{\mathsf{keylen}} $
             \zi \For $i \gets 1$ \To $\id{length}[Pixels]$
             \zi \Do
             Generate $nonce_i$
             \zi $rand = G(k_s,nonce_i)$
             \zi \Comment Choose $rand$ randomly from $[-\delta, \delta]$
             \zi $Pixels[i] \gets Pixels[i] + rand$
             \zi \If $Pixels[i] > 1$
             \zi \Do
             $Pixels[i] = 1$
             \End
             \zi \If $Pixels[i] < 0$
             \zi \Do
             $Pixels[i] = 0$
             \End
             \zi \Comment Clip pixel values to allowed threshold
             \zi $i \gets i+1$
             \End
           \end{codebox}
         \end{minipage}
       }
     }
     \caption{Randomizing image pixels via \our. $Pixels$ represents a vector comprising all pixels of the input image, $G$ denotes a pseudo-random number generator, and~$nonce_i$ is a fresh nonce for every~$i$.}
     \label{fig: pixel randomization}
   \end{figure}

Note that we introduce randomness to the input only while testing and not while training:
this makes our technique particularly lightweight and versatile, as it does not increase training costs and can be applied directly to any pre-trained classifier.
In addition, \our{} invokes the underlying model only once per prediction, without relying on ensembling, which also improves efficiency compared to~\croprescale{} and~\regionbased.
  
\subsection{Analysis: Attacks' Categorization}\label{sec:attack:categorization}

We briefly discuss the attacks considered in our evaluation of randomized defenses (\cf~Section~\ref{sec:eval and results}) in the context of the attacker models from the previous section.

\para{Graybox Attacks.}
Prominent techniques to generate adversarial examples against DL classifiers are the Fast Gradient Sign Method (FGSM)~\cite{Goodfellow2015}, the Basic Iterative Method (BIM)~\cite{DBLP:journals/corr/KurakinGB16}, the Jacobian Saliency Map Approach (JSMA)~\cite{DBLP:journals/corr/KurakinGB16}, the Carlini-Wagner (CW) attacks~\cite{Carlini017}, and DeepFool~\cite{DBLP:conf/cvpr/Moosavi-Dezfooli16}.
These attacks were specifically designed to fool neural networks in a whitebox setting, hence they assume that architecture and parameters are known to the attacker.
Generating adversarial samples according to any of the aforementioned attacks, and then using these samples against an enhanced version of the neural network via some defense mechanism,%
\footnote{Here: \croprescale, \regionbased, or~\our.}
results in a \emph{graybox} attack---because the neural network's internals are available to the attacker, but the defense is not.

\para{Whitebox Attacks.}
Athalye~\etal~\cite{ICML:AthalyeEIK18} proposed the Expectation over Transformation (EOT), a generic method to generate adversarial samples that remain adversarial over a chosen distribution of transformations---in particular over randomized ones.
EOT can handle only differentiable transformations, hence it is not applicable to image-denoising defenses such as~\featuresq.
A second technique, Backward-Pass Differentiable Approximation (BPDA)~\cite{ICML:AthalyeC018}, was introduced to cope with non-differentiable transformations and later employed to defeat, among others, a generalized version of~\featuresq{}~\cite{pixelDiscr}.
We evaluate~\croprescale, \regionbased, and~\our{} against the BPDA and EOT attacks, individually and in combination, appropriately tuned to each defense. Due to exploiting knowledge of the underlying neural-network parameters and of the defense fully, both BPDA and the combination BPDA$+$EOT yield whitebox (i.e., ``fully adaptive'') attacks.

\section{Evaluation and Results}\label{sec:eval and results}

In this section, we evaluate the randomness-based defenses presented in Section~\ref{sec:solution} against graybox and whitebox attacks.
Specifically, we compare~\featuresq{}~\cite{Xu2017} and \our{} by testing them against 11 state-of-the-art graybox attacks, showing that randomness hardens \featuresq{}.
We further evaluate all three randomness-based defenses against the whitebox attacks proposed by Athalye~\etal~\cite{ICML:AthalyeC018}, and explore how increasing the amounts of randomness affects their success.


\subsection{Setup}

\para{Attacks.}
As proposed by Xu~\etal~\cite{Xu2017}, we analyze two variations of each targeted attack: (i) next: targets the class next to the ground truth class modulo number of classes (ii) least-likely (LL): targets the class which the image is least-likely to be classified as.
Specifically, we consider the following attacks on~\featuresq{} and~\our{}: Fast Gradient Sign Method (FGSM)~\cite{Goodfellow2015}, Basic Iterative Method (BIM)~\cite{DBLP:journals/corr/KurakinGB16}, Carlini and Wagner $L^{0}$, $L^{2}$ and $L^{\infty}$ attacks (CW)~\cite{Carlini017} (Next \& LL), DeepFool~\cite{DBLP:conf/cvpr/Moosavi-Dezfooli16}, Jacobian Saliency Map Approach (JSMA)~\cite{DBLP:conf/eurosp/PapernotMJFCS16} (Next \& LL).
We further evaluate \our{}, \croprescale{}, and~\regionbased, against fully adaptive attacks  (BPDA and EOT) proposed by Athalye~\etal~\cite{ICML:AthalyeC018}.

\para{Datasets.} We use ImageNet, CIFAR-10, and MNIST  datasets to conduct our experiments.
The ImageNet dataset contains 1.2 million images for training and 50\,000 images for validation. They are of various sizes and hand-labeled with 1000 classes. The images are preprocessed to $224 \times 224$ pixels and encoded with 24-bit color per pixel.
CIFAR-10 is a dataset of $32 \times 32$ pixel images with 24-bit color per pixel (three color channels per pixel) and 10 classes.
MNIST is a dataset of hand-written digits (0-9) encoded as 8-bits grayscale images of size $28 \times 28$ pixels (one color channel per pixel).

\para{Target Models.} We tested the aforementioned attacks on the same pre-trained models as in Feature Squeezing~\cite{Xu2017}. Namely,
we use MobileNet~\cite{DBLP:journals/corr/HowardZCKWWAA17} for ImageNet,
a 7-layer CNN for MNIST\footnote{\url{https://github.com/carlini/nn_robust_attacks/}},
and a DenseNet model for CIFAR-10\footnote{\url{https://github.com/titu1994/DenseNet/}}~\cite{DBLP:conf/cvpr/HuangLMW17}.
These models achieve a top-1 accuracy of 99.43\%, 94.84\%, and 68.36\%, respectively.
The prediction performance of these models is at par with best models\footnote{\url{http://rodrigob.github.io/are_we_there_yet/build/classification_datasets_results.html}}.
To study the effects of introducing randomness, we use the same data samples as used by Xu~\etal~\cite{Xu2017}. We use 100 adversarial samples for each of the 11 attacks for all datasets. Each color channel of the pixel is normalized to be in the range $[0,1]$. We use 10\,000 legitimate samples for CIFAR-10 and MNIST, and 200 samples for ImageNet (due to high computation cost) to study the effect of adding randomness to the defenses.

\para{Experimental Setup.} We evaluate the efficacy of the 3 defenses proposed by Xu~\etal~\cite{Xu2017}---bit depth reduction, median smoothing and non-local smoothing, when combined with randomness. We study each of the 11 attacks against the 3 defenses with varying parameters for 3 datasets. The experiment is repeated 200 times for each randomness level to compute the statistics. In our evaluation, the accuracy over adversarial samples is averaged over 200 runs. 
We note that for the deterministic case when no randomness is added ($\delta = 0$) the results do not change.

 \para{Implementation.} We implemented \our{} in Python and executed it using CPython. Namely, we adapted the open source code released by Xu~\etal\footnote{\url{https://github.com/mzweilin/EvadeML-Zoo}} to implement our solution. We use a machine with 3.5 GHz processor and 32 GB RAM for our experiments.


\begin{figure*}[!tb]
  \vspace{-1em}
  \centering
  {\includegraphics[width=\linewidth]{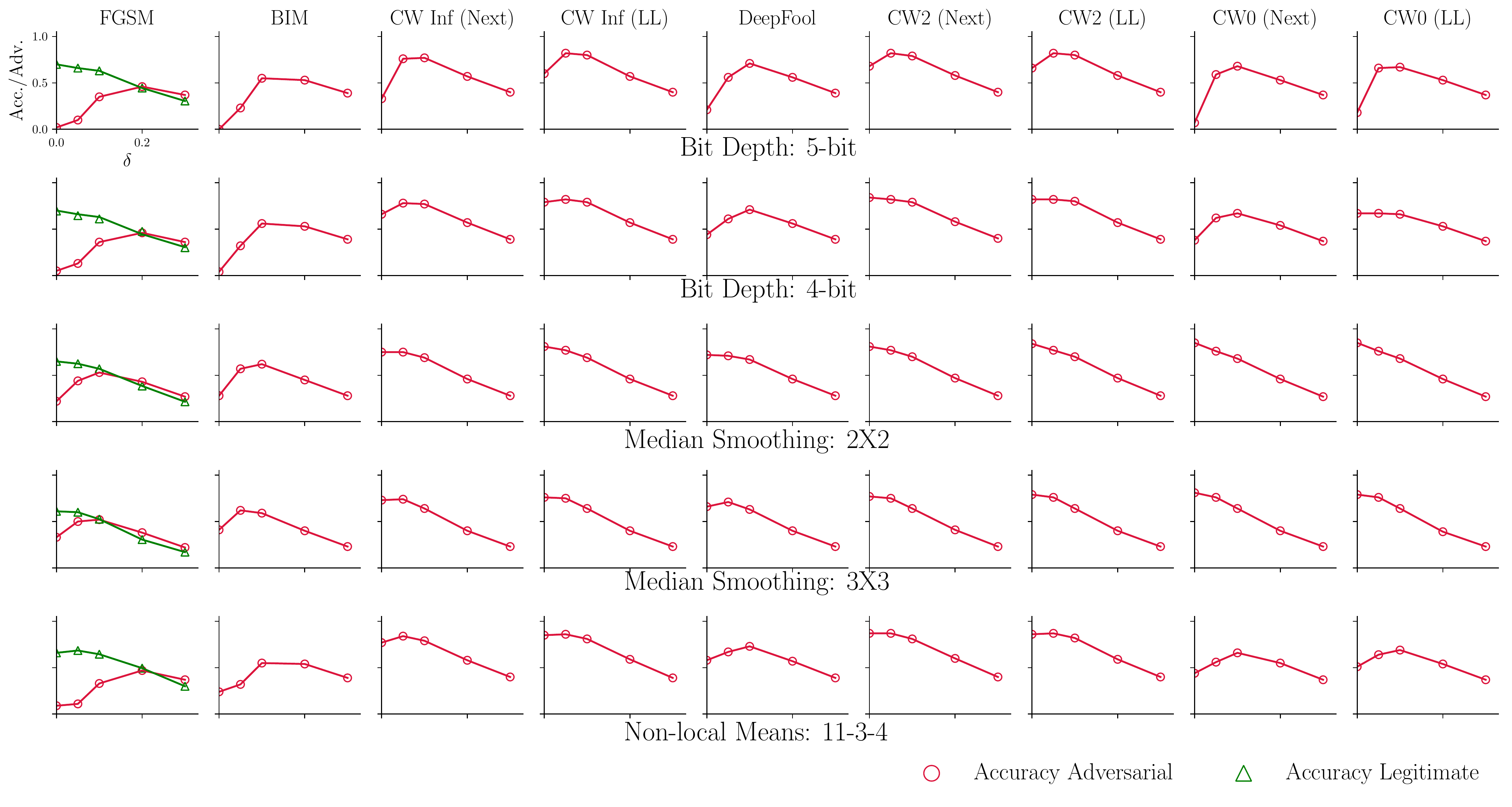}}
  \vspace{-2em}
  \caption{ImageNet: Behavior of accuracy for magnitudes of randomness $\delta = [0, 0.05, 0.1, 0.2, 0.3]$. 
    We also plot the accuracy of the model for legitimate samples as $\delta$ increases (shown once for each defense as the curve does not change).}
  \label{fig:imagenet advantage and accuracy vs rand}
  \vspace{-1em}
\end{figure*}

\subsection{Graybox Adversaries}\label{sec:results}

This section presents the results of our evaluation of \our{}, compared to the deterministic~\featuresq{}, against graybox attacks.

\para{Choosing $\delta$.} Choosing the randomness magnitude $\delta$ 
appropriately is vital to designing an effective defense. The choice depends upon the nature of dataset and defense used. As we show in the paragraphs ahead the behavior of defenses can vary for grayscale and color images. We study these variations extensively by running experiments for changing $\delta$. We present our evaluation of ImageNet, CIFAR-10 and MNIST datasets next.

\para{ImageNet.}
Figure~\ref{fig:imagenet advantage and accuracy vs rand} shows the behavior of accuracy of the classifier for both adversarial and legitimate samples as input randomization ($\delta$) is increased. The accuracy decreases as $\delta$ is increased. Squeezing via reducing bit-depth shows a drastic drop in accuracy beyond a certain randomness for most attacks; as $\delta$ increases, a larger fraction of pixels breach the quantization threshold which results in them being flipped to 0 or 1 despite being very distant earlier, this results in accuracy dropping sharply. The CW0 adversarial samples show an improvement in accuracy for high $\delta$, this is due to large $L^{0}$ perturbations being undone due to noise. Median smoothing methods are less affected by large randomness values due to the randomness being averaged out. Hence we observe a gradual decline in accuracy with increasing randomness. Note that CIFAR-10 has color and each pixel has three color channels each of which are normalized to $[0,1]$. We introduce randomness individually to each channel for each pixel, hence the effect of randomness becomes significant even at low $\delta$ values.

We conclude that an appropriately chosen~$\delta$ provides desirable security properties, we found that $\delta = 0.1$ provides the best trade-off between accuracy on legitimate and adversarial samples. We present the accuracy values in Table~\ref{tab:accuracy feature squeezing random}. We note that the accuracy decreases slightly over legitimate samples. For comparison, Table~\ref{tab:accuracy feature squeezing} shows the accuracy of defenses when no randomness is added~\cite{Xu2017}.

We further compare the robustness of~\our{} with that of the other two randomized defenses, \croprescale{} and \regionbased.
As advocated in~\cite{ICLR:GuoRCM18}, \croprescale{} achieves an accuracy of 45-65\% against FGSM, while \regionbased{} and \our{} achieve an accuracy of 34.7\% and 53.44\% (with median smoothing), respectively.
Similarly, for CW2 Next (Next and LL), \croprescale{} achieves an accuracy of~40-65\% (adapted from~\cite{ICLR:GuoRCM18}), while \regionbased{} and \our{} achieve an accuracy of~79\% and~70\% (with median smoothing), respectively.
This shows that online randomization defenses offer decent robustness to gray-box attacks even when compared to offline defenses.


\begin{figure*}[!htb]
	\centering
	{\includegraphics[width=\linewidth]{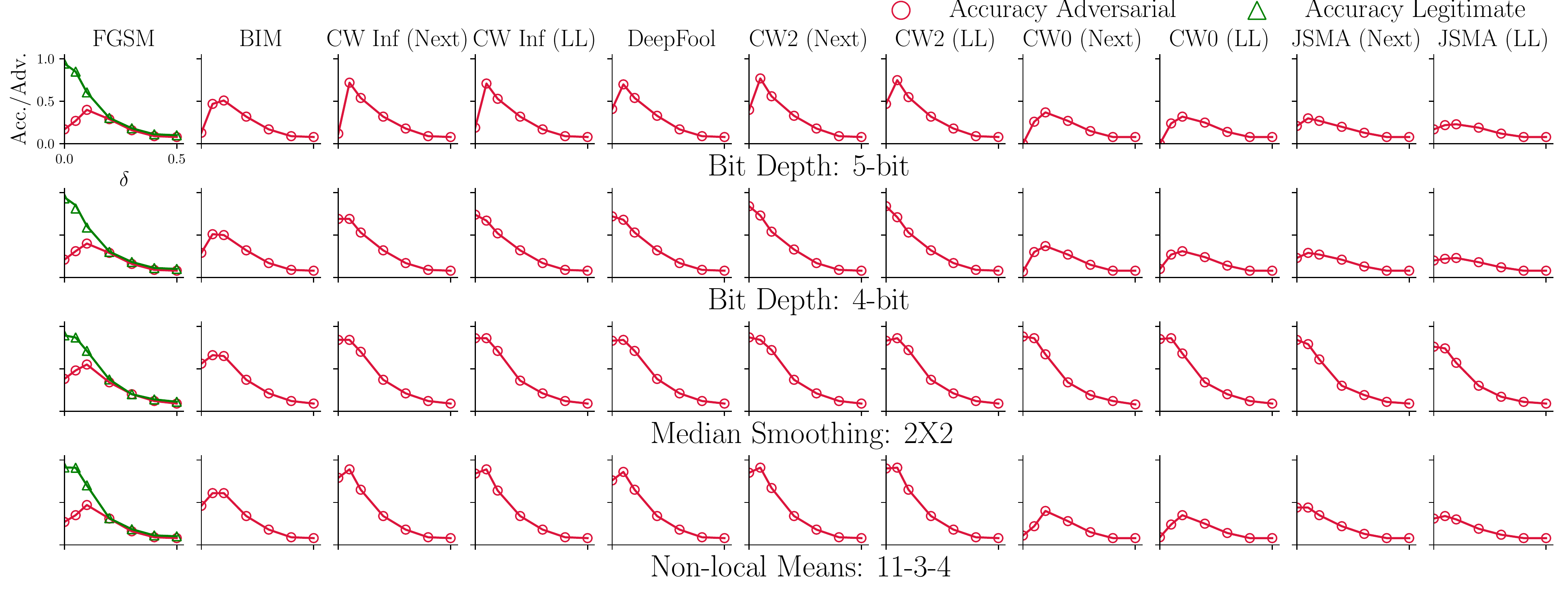}}
	\caption{CIFAR-10: Behavior of accuracy for magnitudes of randomness $\delta = [0, 0.05, 0.1, 0.2, 0.3, 0.4, 0.5]$.	We also plot the accuracy of the model for legitimate samples as $\delta$ increases (shown once for each defense as the curve does not change).}
	\label{fig:cifar advantage and accuracy vs rand}
\end{figure*}


\para{CIFAR-10.}
The results for running \our{} within the CIFAR-10 dataset are similar to ImageNet: adding randomness helps make misclassified samples unpredictable. Figure~\ref{fig:cifar advantage and accuracy vs rand} 
shows that accuracy over both legitimate and adversarial samples drops sharply on increasing $\delta$. Identical to our evaluation in the ImageNet dataset, we note that \our{} introduces small randomness for each color channel. The accuracy over adversarial samples improves significantly at $\delta = 0.05$ for almost all defenses with just a small drop over legitimate samples (\cf~Tables~\ref{tab:accuracy feature squeezing} \&~\ref{tab:accuracy feature squeezing random}). Large values of $\delta$ make the classifier unusable as accuracy drops.

\begin{figure*}[!htb]
   \centering
   {\includegraphics[width=\linewidth]{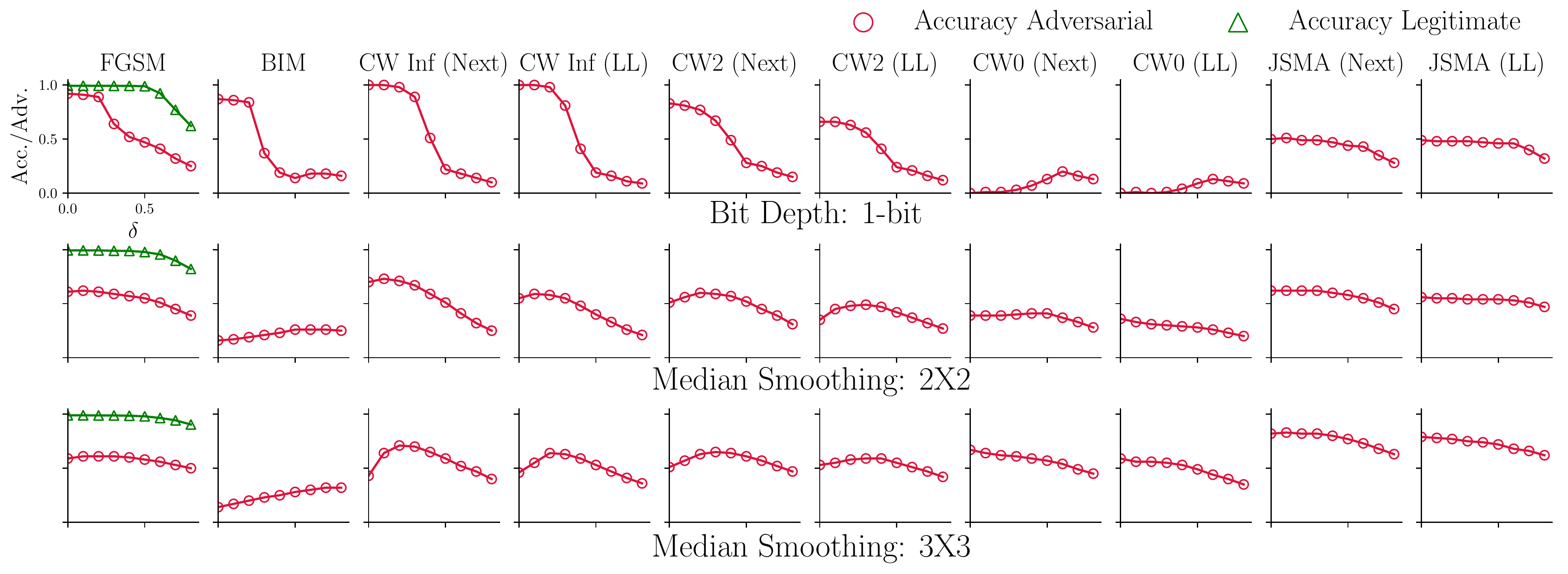}}
   \caption{MNIST: Behavior of accuracy  
   	for $\delta = [0, 0.1, 0.2, 0.3, 0.4, 0.5, 0.6, 0.7, 0.8]$. 
    We also plot the accuracy of the model for legitimate samples as $\delta$ increases (shown once for each defense as the curve does not change).}
   \label{fig:mnist advantage and accuracy vs rand}
 \end{figure*}

\para{MNIST.}
As seen in Figure~\ref{fig:mnist advantage and accuracy vs rand}, 
the results of running \our{} with the MNIST dataset are similar to ImageNet and CIFAR-10. We present the accuracy values in Tables~\ref{tab:accuracy feature squeezing} \&~\ref{tab:accuracy feature squeezing random}. The behavior of accuracy for the MNIST dataset is similar to that of ImageNet and CIFAR-10 (\cf~Figures~\ref{fig:cifar advantage and accuracy vs rand}).


\begin{figure*}[!htb]
   \centering
   {\includegraphics[width=\linewidth]{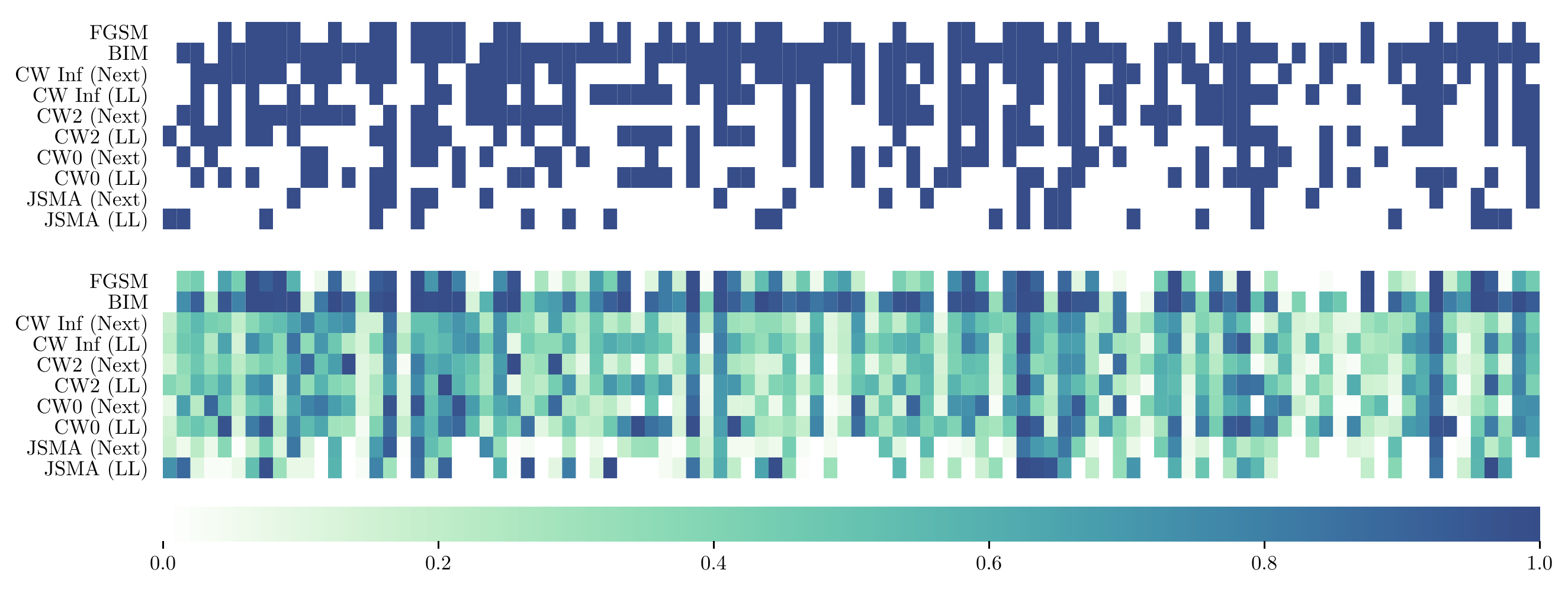}}
   \caption{MNIST: Unpredictability of errors for Median Smoothing ($3 \times 3$) defense, without randomness (top figure, $\delta = 0$) and with randomness (bottom figure, $\delta = 0.5$)}
   \label{fig:mnist unpredictable errors}
 \end{figure*}

Figure~\ref{fig:mnist unpredictable errors} 
shows the probabilities of the prediction errors when used with $3 \times 3$ median smoothing with and without randomness. 
Each row represents a specific attack strategy~$\adv$; the $x$-axis represents the adversarial sample set~$\Advset$ (in particular $\size{\Advset} = 100$); the color intensity of each cell indicates the estimated probability (over the classifier's randomness) that the corresponding adversarial sample succeeds (note, this probability is binary for a deterministic classifier). However, when randomness is introduced the error probabilities spread out for a large number of samples as they are no longer deterministic. For each of the considered attacks~$\adv$, the empirical error over~$\Advset$ can be computed by summing up the probabilities that each adversarial sample in~$\Advset$ succeeds, \ie, $\err_{\Advset}(\classifier_\$)   =  \sum_{i=1}^{100} \prob{\classifier_\$(x_i) \neq f(x_i)}$. For ease of presentation, we only present the results applied to one defense in order to to demonstrate the effect of randomness, the results for other defenses are similar.


\begin{table*}[!tb]
  \caption{Accuracy of original Feature Squeezing defenses without randomness ($\delta = 0$) over adversarial samples (\%). Xu~\etal~\cite{Xu2017} omit DeepFool on MNIST as the adversarial samples generated appear unrecognizable to humans; non-local smoothing is not applied to MNIST as it is hard to find similar patches on such images for smoothing a center patch. JSMA is omitted for ImageNet due to large memory requirement.}
\begin{center}
\resizebox{\linewidth}{!}{%
\begin{tabular}{@{}c|c|c|c|c|c|c|c|c|c|c|c|c|c|c|c@{}}
\toprule
\multirow{3}{*}{\textbf{Dataset}}  & \multicolumn{2}{c|}{\textbf{Squeezer}}                 & \multicolumn{4}{c|}{\textbf{$L^{\infty}$ Attacks}}     & \multicolumn{3}{c|}{\textbf{$L^{2}$ Attacks}}             & \multicolumn{4}{c|}{\textbf{$L^{0}$ Attacks}}            & \multirow{3}{*}{\textbf{All Attacks}} & \multirow{3}{*}{\textbf{Legitimate}} \\ \cmidrule(lr){2-14}
                          & \multirow{2}{*}{Name}             & \multirow{2}{*}{Parameters} & \multirow{2}{*}{FGSM} & \multirow{2}{*}{BIM} & \multicolumn{2}{c|}{$CW_\infty$} & \multirow{2}{*}{DeepFool}        & \multicolumn{2}{c|}{$CW_2$} & \multicolumn{2}{c|}{$CW_0$} & \multicolumn{2}{c|}{JSMA} &                              &                             \\ \cmidrule(lr){6-7} \cmidrule(lr){9-14}
                          &                                   &                             &                       &                      & Next        & LL         &                           & Next         & LL        & Next         & LL        & Next         & LL         &                              &                             \\ \midrule
\multirow{4}{*}{MNIST}    & \multicolumn{2}{c|}{None}                                       & 54                    & 9                    & 0           & 0          & -                         & 0            & 0         & 0            & 0         & 27           & 40         & 13.00                        & 99.43                       \\ \cmidrule(l){2-16}
                          & Bit Depth                         & 1-bit                       & 92                    & 87                   & 100         & 100        & -                         & 83           & 66        & 0            & 0         & 50           & 49         & 62.70                        & 99.33                       \\ \cmidrule(l){2-16}
                          & \multirow{2}{*}{Median Smoothing} & $2 \times 2$                & 61                    & 16                   & 70          & 55         & -                         & 51           & 35        & 39           & 36        & 62           & 56         & 48.10                        & 99.28                       \\ \cmidrule(l){3-16}
                          &                                   & $3 \times 3$                & 59                    & 14                   & 43          & 46         & -                         & 51           & 53        & 67           & 59        & 82           & 79         & 55.30                        & 98.95                       \\ \midrule\midrule
\multirow{5}{*}{CIFAR-10} & \multicolumn{2}{c|}{None}                                       & 15                    & 8                    & 0           & 0          & 2                         & 0            & 0         & 0            & 0         & 0            & 0          & 2.27                         & 94.84                       \\ \cmidrule(l){2-16}
                          & \multirow{2}{*}{Bit Depth}        & 5-bit                       & 17                    & 13                   & 12          & 19         & 40                        & 40           & 47        & 0            & 0         & 21           & 17         & 20.55                        & 94.55                       \\ \cmidrule(l){3-16}
                          &                                   & 4-bit                       & 21                    & 29                   & 69          & 74         & 72                        & 84           & 84        & 7            & 10        & 23           & 20         & 44.82                        & 93.11                       \\ \cmidrule(l){2-16}
                          & Median Smoothing                  & $2 \times 2$                & 38                    & 56                   & 84          & 86         & 83                        & 87           & 83        & 88           & 85        & 84           & 76         & 77.27                        & 89.29                       \\ \cmidrule(l){2-16}
                          & Non-local Means                   & 11-3-4                      & 27                    & 46                   & 80          & 84         & 76                        & 84           & 88        & 11           & 11        & 44           & 32         & 53.00                        & 91.18                       \\ \midrule\midrule
\multirow{5}{*}{ImageNet} & \multicolumn{2}{c|}{None}                                       & 1                     & 0                    & 0           & 0          & 11                        & 10           & 3         & 0            & 0         & -            & -          & 2.78                         & 69.70                       \\ \cmidrule(l){2-16}
                          & \multirow{2}{*}{Bit Depth}        & 5-bit                       & 2                     & 0                    & 33          & 60         & 21                        & 68           & 66        & 7            & 18        & -            & -          & 30.56                        & 69.40                       \\ \cmidrule(l){3-16}
                          &                                   & 4-bit                       & 5                     & 4                    & 66          & 79         & 44                        & 84           & 82        & 38           & 67        & -            & -          & 52.11                        & 68.00                       \\ \cmidrule(l){2-16}
                          & \multirow{2}{*}{Median Smoothing} & $2 \times 2$                & 22                    & 28                   & 75          & 81         & 72                        & 81           & 84        & 85           & 85        & -            & -          & 68.11                        & 65.40                       \\ \cmidrule(l){3-16}
                          &                                   & $3 \times 3$                & 33                    & 41                   & 73          & 76         & 66                        & 77           & 79        & 81           & 79        & -            & -          & 67.22                        & 62.10                       \\ \cmidrule(l){2-16}
                          & Non-local Means                   & 11-3-4                      & 10                    & 25                   & 77          & 82         & 57                        & 87           & 86        & 43           & 47        & -            & -          & 57.11                        & 65.40                       \\ \bottomrule
\end{tabular}}
\end{center}
\label{tab:accuracy feature squeezing}
\end{table*}


\begin{table*}[tb]
\caption{Accuracy of \our{} (\%). Bold values indicate an improvement over corresponding values in Table~\ref{tab:accuracy feature squeezing}.}
\begin{center}
\resizebox{\linewidth}{!}{%
\begin{tabular}{@{}c|c|c|c|c|c|c|c|c|c|c|c|c|c|c|c@{}}
\toprule
\multirow{3}{*}{\textbf{Dataset}}  & \multicolumn{2}{c|}{\textbf{Squeezer}}                   & \multicolumn{4}{c|}{\textbf{$L^{\infty}$ Attacks}}     & \multicolumn{3}{c|}{\textbf{$L^{2}$ Attacks}}             & \multicolumn{4}{c|}{\textbf{$L^{0}$ Attacks}}            & \multirow{3}{*}{\textbf{All Attacks}} & \multirow{3}{*}{\textbf{Legitimate}} \\ \cmidrule(lr){2-14}
                          & \multirow{2}{*}{Name}                                             & \multirow{2}{*}{Parameters} & \multirow{2}{*}{FGSM} & \multirow{2}{*}{BIM} & \multicolumn{2}{c|}{$CW_\infty$} & \multirow{2}{*}{DeepFool}        & \multicolumn{2}{c|}{$CW_2$} & \multicolumn{2}{c|}{$CW_0$} & \multicolumn{2}{c|}{JSMA} &                              &                             \\ \cmidrule(lr){6-7} \cmidrule(lr){9-14}
                          &                                                                   &                             &                       &                      & Next           & LL                &                           & Next                  & LL                & Next            & LL              & Next           & LL                 &                              &                             \\ \midrule
\multirow{4}{*}{MNIST}    & \makecell{Bit Depth\\($\delta = 0.2$)}                            & 1-bit                       & 88.98                 & 84.06                & 98.37          & 98.26             & -                         & 77.44                 & 63.19             & \textbf{1.39}   & \textbf{0.42}   & 49.26          & 48.41              & 60.98                        & 99.31                       \\ \cmidrule(l){2-16}
                          & \multirow{2}{*}{\makecell{Median Smoothing\\($\delta = 0.5$)}}    & $2 \times 2$                & 54.78                 & \textbf{25.53}       & 50.70          & 40.22             & -                         & \textbf{52.20}        & \textbf{42.21}    & \textbf{40.65}  & 28.27           & 58.15          & 54.05              & 44.68                        & 97.54                       \\ \cmidrule(l){3-16}
                          &                                                                   & $3 \times 3$                & 58.03                 & \textbf{27.61}       & \textbf{58.53} & \textbf{52.87}    & -                         & \textbf{61.27}        & \textbf{55.40}    & 56.86           & 48.78           & 77.23          & 71.67              & \textbf{56.83}               & 97.99                       \\ \midrule\midrule
\multirow{5}{*}{CIFAR-10} & \multirow{2}{*}{\makecell{Bit Depth\\($\delta = 0.05$)}}          & 5-bit                       & \textbf{27.33}        & \textbf{47.27}       & \textbf{71.75} & \textbf{70.89}    & \textbf{69.91}            & \textbf{76.90}        & \textbf{75.06}    & \textbf{26.21}  & \textbf{23.89}  & \textbf{29.84} & \textbf{22.06}     & \textbf{49.19}               & 85.03                       \\ \cmidrule(l){3-16}
                          &                                                                   & 4-bit                       & \textbf{30.70}        & \textbf{51.34}       & 68.89          & 67.29             & 67.55                     & 72.95                 & 71.08             & \textbf{30.25}  & \textbf{26.60}  & \textbf{29.36} & \textbf{22.14}     & \textbf{48.92}               & 81.27                       \\ \cmidrule(l){2-16}
                          & \makecell{Median Smoothing\\($\delta = 0.05$)}                    & $2 \times 2$                & \textbf{47.66}        & \textbf{66.23}       & 83.59          & 85.90             & \textbf{83.70}            & 84.45                 & \textbf{86.50}    & 85.61           & \textbf{86.11}  & 78.54          & 73.62              & \textbf{78.36}               & 86.80                       \\ \cmidrule(l){2-16}
                          & \makecell{Non-local Means\\($\delta = 0.05$)}                     & 11-3-4                      & \textbf{35.27}        & \textbf{61.19}       & \textbf{88.59} & \textbf{89.39}    & \textbf{86.31}            & \textbf{91.01}        & \textbf{90.52}    & \textbf{22.08}  & \textbf{24.21}  & \textbf{44.05} & \textbf{33.80}     & \textbf{60.58}               & 90.90                       \\ \midrule\midrule
\multirow{5}{*}{ImageNet} & \multirow{2}{*}{\makecell{Bit Depth\\($\delta = 0.1$)}}           & 5-bit                       & \textbf{34.86}        & \textbf{54.97}       & \textbf{76.97} & \textbf{79.50}    & \textbf{70.73}            & \textbf{79.45}        & \textbf{80.31}    & \textbf{67.52}  & \textbf{67.43}  & -              & -                  & \textbf{67.97}               & 63.00                       \\ \cmidrule(l){3-16}
                          &                                                                   & 4-bit                       & \textbf{35.63}        & \textbf{56.11}       & \textbf{76.66} & 78.92             & \textbf{70.91}            & 79.11                 & 79.69             & \textbf{67.40}  & 66.48           & -              & -                  & \textbf{67.88}               & 61.00                       \\ \cmidrule(l){2-16}
                          & \multirow{2}{*}{\makecell{Median Smoothing\\($\delta = 0.1$)}}    & $2 \times 2$                & \textbf{53.44}        & \textbf{62.49}       & 69.11          & 69.09             & 67.22                     & 70.13                 & 69.97             & 68.47           & 68.20           & -              & -                  & 66.46                        & 57.00                       \\ \cmidrule(l){3-16}
                          &                                                                   & $3 \times 3$                & \textbf{52.23}        & \textbf{58.88}       & 63.76          & 63.88             & 62.87                     & 64.46                 & 63.91             & 63.76           & 64.22           & -              & -                  & 62.00                        & 52.50                       \\ \cmidrule(l){2-16}
                          & \makecell{Non-local Means\\($\delta = 0.1$)}                      & 11-3-4                      & \textbf{32.53}        & \textbf{55.29}       & \textbf{78.97} & 81.28             & \textbf{72.92}            & 81.11                 & 81.52             & \textbf{66.02}  & \textbf{68.71}  & -              & -                  & \textbf{68.71}               & 64.50                       \\ \cmidrule(l){2-16}
                          & \makecell{Pure Randomness}                                        & $\delta = 0.1$              & 34.70                 & 56.33                & 76.91          & 79.58             & 68.84                     & 79.05                 & 80.41             & 67.06           & 67.5            & -              & -                  & 67.82                        & 64                          \\ \bottomrule
\end{tabular}}
\end{center}
\label{tab:accuracy feature squeezing random}
\end{table*}


\para{Interpretation of Results.} Increase in the magnitude of randomness drives the accuracy of the classifier over legitimate samples towards 10\% as classification becomes akin to guessing (for classification over 10 classes as in MNIST and CIFAR-10). We made a deliberate choice to clip the pixel values when they go outside the allowed bounds of $[0,1]$ rather than wrapping around. A value of $\delta = 1$ and wrapping around the pixel values when they go out of bounds produces a truly random pixel, and hence the image. We found that at this level of randomness, accuracy over legitimate samples becomes close to 10\%. Even lower values of $\delta$ produce a sharp drop in accuracy over legitimate samples, hence we choose to clip the values when they go out of bounds. 

The primary motivation for our design of~\our{} is to perturb the pixels in a manner which subsumes the adversarial perturbation, and to which the adversary cannot adapt while keeping the usefulness of the classifier intact. The optimum magnitude of randomness~$\delta$ to be used is contingent on the defense used. High values of~$\delta$ have strong effect on the accuracy when used in conjunction with bit depth reduction, as it could change the value of a pixel drastically if the bit depth is low. In contrast, methods like local and non-local smoothing are much more resilient to high~$\delta$, as they average out the noise from sections of images. The noise that we add, being additive, is filtered out. 

Note that we want to use the largest value of~$\delta$ possible so as to subsume the adversarial perturbations, while still maintaining high accuracy. Not all defenses are equally potent for all attacks and datasets, therefore the randomness magnitude~$\delta$ must be carefully chosen. \our{} can mitigate the limitation of a weak defense to some extent, as seen in the case of CIFAR-10 bit depth (5-bit) defense where the accuracy over adversarial samples increases by almost 2.5 times. However, efficacy of the defense is critical to have success in general.


\subsection{Whitebox Adversaries}\label{sec:adaptive adversaries}


\begin{figure*}[!htb]
  \centering
  \begin{minipage}{.49\linewidth}
    \includegraphics[width=\linewidth]{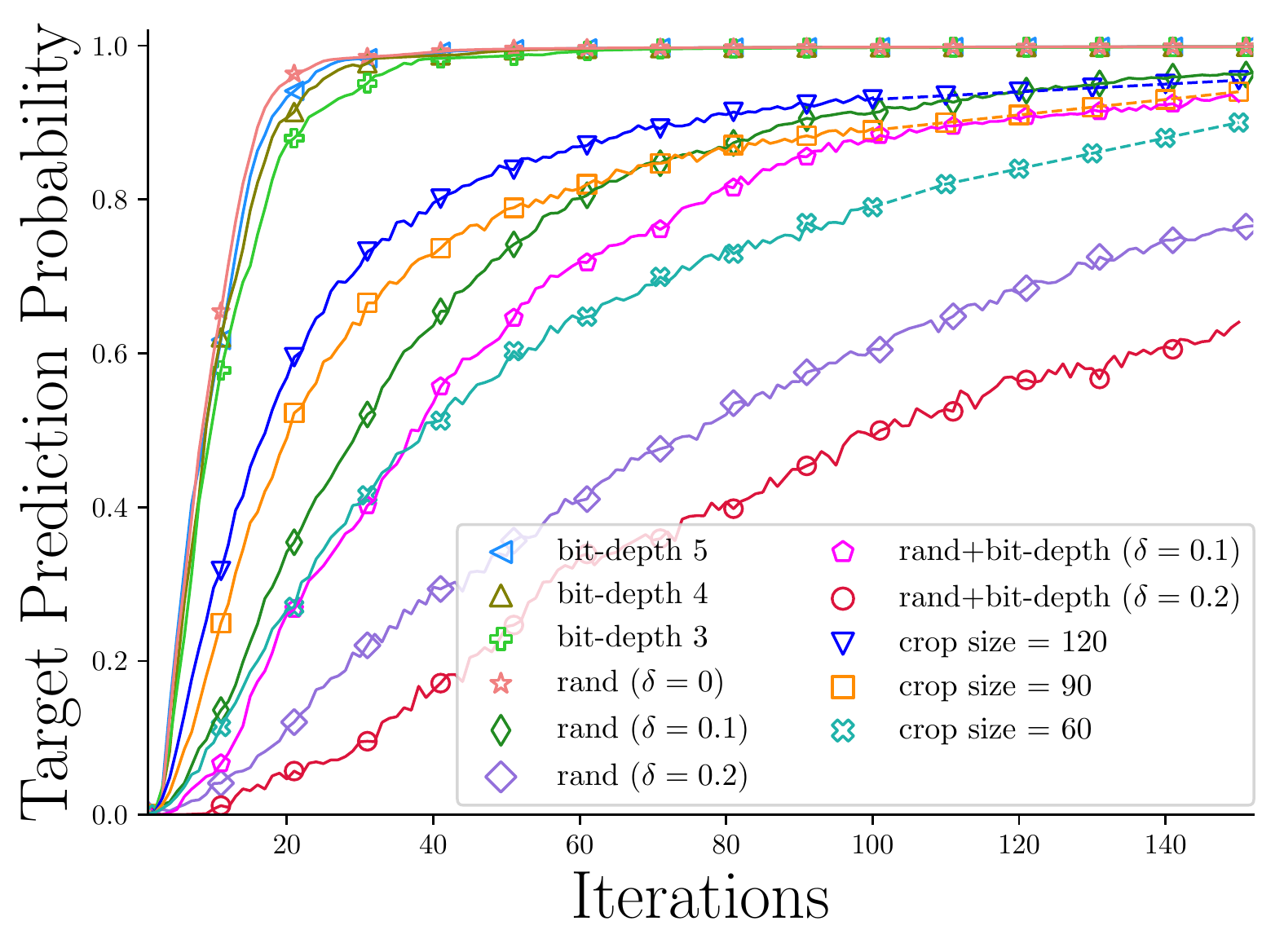}
  \end{minipage}
  \begin{minipage}{.49\linewidth}
    \includegraphics[width=\linewidth]{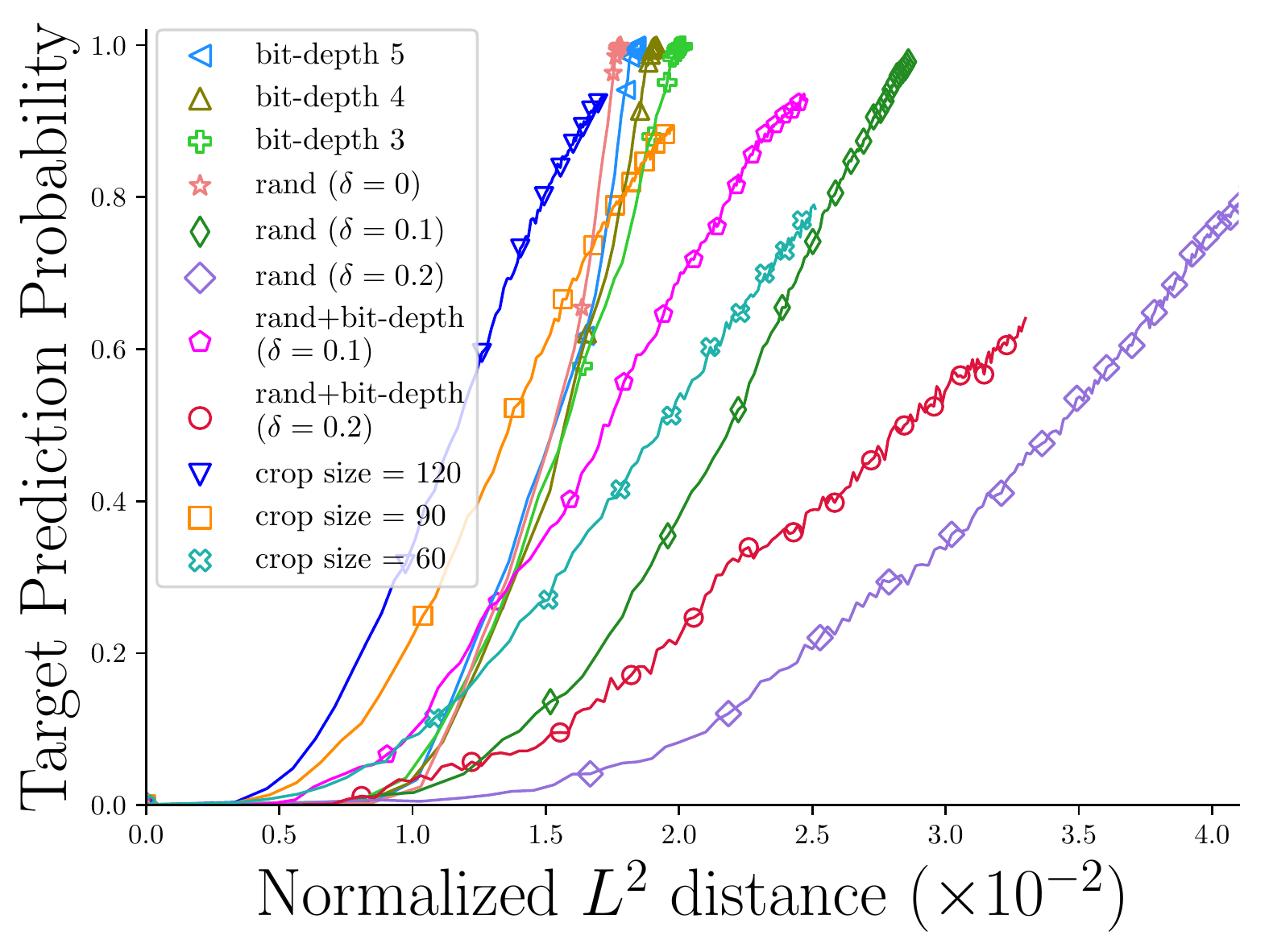}
  \end{minipage}
  \vspace{-1em}
  \caption{Increase in target prediction probability and distortion (in normalized $L^2$ distance) as more confident adversarial samples are synthesized (dashed lines show extrapolated values).}
  \vspace{-1em}
  \label{fig:prob and L2 distance}
 \end{figure*}
 

In this section, we study whether randomness influences the success of whitebox adversaries.
We evaluate the BPDA and EOT attacks proposed by Athalye~\etal~\cite{ICML:AthalyeC018} against \croprescale{}~\cite{ICLR:GuoRCM18}, \regionbased{}~\cite{ACSAC:CaoG17},%
\footnote{Our implementation of~\regionbased{} slightly differs from the originally proposed version~\cite{ACSAC:CaoG17}, namely we use the same randomization strategy as~\our{} (which is equivalent to sampling at random from a hypercube, \cf~Section~\ref{sec:solution:our}) and classify one sample instead of ensembling. Due to averaging over~100 images, ensembling would lead to the same results as adversary can chose to defeat a majority of samples.}
and~\our{} with bit-depth reduction as squeezing function.
Beyond tuning the attacks to each defensive technique, we chose the attack strategies that are best suited for each of the defenses, as outlined in~\cite{ICML:AthalyeC018}.
Specifically, we consider:
a pure instantiation of EOT against~\croprescale, as this defense applies differentiable transformations;
the BPDA attack against~\regionbased, and a combination of BPDA and EOT against~\our, so that both the non-differentiability of squeezing and the input randomization are taken into account when generating adversarial samples.
For completeness, we also ran the attacks against the deterministic bit-depth reduction.

We evaluated the defenses against the aforementioned attacks on the same set of 100 images, selected at random from the ImageNet dataset. Each image was assigned a target class at random.
Figure~\ref{fig:prob and L2 distance} summarizes the results.

Recall that every attack iteration aims to generate adversarial samples with higher confidence compared to the previous iteration.
Correspondingly, in the leftmost plot we depict the (average) target prediction probability against the number of iterations.
We see that the prediction probability starts at~0 for all defenses, and approaches~1 as the number of iterations increases.
Notice that the prediction probability quickly reaches~1 for defenses which have no randomness whereas it grows gradually for randomness-based defenses.
For instance, when randomness is not applied, less than~20 iterations are sufficient to achieve a prediction probability of~0.8.
In contrast, for randomized defenses 20 iterations lead to a prediction probability of only~0.6 in the best case (i.e., \croprescale{} with crop size~120), and of less than~0.05 in the case of the seemingly most robust technique (i.e., \our{} for~$\delta = 0.2$).
Comparing among the randomness-based defenses, we observe that for a prediction probability of~$0.6$, the attacks requires about
20, 35, or~50 iterations for~\croprescale{} depending on the crop size,
35 or 100 iterations in the case of~\regionbased{} for randomness magnitudes~$\delta = 0.1$ and $\delta =0.2$, respectively,
and 45 or 140 iterations against~\our{} for the same values of~$\delta$, indicating that both online defenses outperform~\croprescale{}.

We also illustrate how the target prediction probability varies with the distortion, measured as normalized $L^2$ distance (see the rightmost plot in Figure~\ref{fig:prob and L2 distance}).
Again, we see that randomness-based defenses are more robust than their deterministic counterparts, as they force the attacker to introduce larger perturbation.
Namely, for all deterministic defenses, high-confidence adversarial samples can be generated with a distortion below~0.02, 
while larger perturbations are necessary to defeat randomized defenses.
In particular, high-confidence adversarial samples against~\regionbased{} and~\our{} require perturbations with~$L^2$ distance above~0.025 (for $\delta = 0.1$), and above~0.035 (for $\delta = 0.2$), respectively, with~\regionbased{} presenting slightly higher robustness than~\our{} according to this metric. 
Our results support the intuition that introducing unpredictability to the classification process makes it computationally expensive to find adversarial perturbations which are successful regardless of the randomness.

\section{Related Work}
\label{sec:relatedwork}

There has been massive effort to make ML models robust to adversarial samples, leading to a huge number of defenses proposed in the last few years.%
\footnote{\url{https://nicholas.carlini.com/writing/2019/all-adversarial-example-papers.html}} 
Here, we discuss only a selection of these proposals which we find representative of general defensive principles. 
Following widely-adopted nomenclature, we differentiate between \emph{certified} and \emph{heuristic} defenses to distinguish between proposals that come with provable guarantees and those which do not.

\para{Heuristic defenses.}
An early proposal is \emph{defensive distillation}~\cite{Papernot2016,PapernotM17}, which extends the deep-learning concept of distillation to the adversarial setting, aiming to extract knowledge from a given neural-network architecture to improve its own resilience to gradient-based attacks.
This proposal has been shown ineffective~\cite{CORR:PapernotM17}.

A broad class of defenses attempts to \emph{detect} whether a given input is adversarial. Early methods derive statistical properties from large datasets of legitimate, respectively, adversarial samples,
and then inspect these properties to discern the adversarial nature of new and unknown samples~\cite{GrosseMP0M17,2017arXiv170300410F}.
Even though these techniques where shown to be quite robust, they are computationally expensive and require large datasets for the reliability of statistical results. 
A more efficient approach is to train a detector to specifically learn adversarial samples, as done by MagNet~\cite{MengC17}.
MagNet relies on the assumption that natural data lie on a manifold of significantly smaller dimension than the whole input space, while adversarial samples fall outside the manifold. Based on this, Magnet employs a detector which deems samples far from the manifold as adversarial.
A different kind of detector is instantiated by the feature squeezing defense by Xu~\etal~\cite{Xu2017} (cf.~Section~\ref{sec:background:defenses}), which uses the discrepancy of output predictions between original samples and their squeezed versions to detect adversarial manipulations.

We note that~\our{}, while using squeezing routines, does not aim at detecting adversarial samples, rather at making it harder for the adversary to generate successful perturbations.
Instead, our proposal can be seen as an instantiation of so-called \emph{input transformation}, which applies a pre-processing step to the input in order to reduce the sensitivity of the model to small changes in input---with the hope of making the classifier more robust to adversarial perturbations~\cite{ICLR:GuoRCM18}.

The most robust among all heuristic defenses to date appears to be \emph{adversarial training}, introduced in~\cite{Goodfellow2015} and later extended in several works~\cite{Kurakin2017,ICLR:MadryMSTV18}. Adversarial training essentially finds adversarial samples by running known attacks, and adds those samples to the training set so that the model learns to correctly classify certain adversarial inputs.
Madry~\etal~\cite{ICLR:MadryMSTV18} propose a generic training methodology targeting robustness against all low-distortion adversarial samples, i.e., samples generated by applying small perturbations to clean inputs.
This methodology is based on the idea that, if the training set contains sufficiently representative adversarial samples, the resulting classifier will be able to withstand \emph{all} low-distortion attacks.
Based on this, the authors heuristically generate ``sufficiently representative'' adversarial samples using projected gradient descent (PDG), an attack strategy which generalizes first-order attacks.
The resulting trained networks, based on MNIST and CIFAR datasets respectively, achieve different levels of robustness against FGSM, PGD, and CW attacks. 
Although various works demonstrated the feasibility of this technique, adversarial training requires a large number of samples that are expensive to generate.
In addition, it cannot resist unknown attacks.

Except for adversarial training, all the aforementioned techniques were shown, in a way or another, vulnerable to adaptive attacks.

\para{Certified defenses.}
The idea of ensuring robustness to \emph{all} attacks within a certain class has been investigated further, fostering a line of work aimed at developing \emph{certified defenses}~\cite{NIPS:HeinA17,ICML:WongK18,raghunathan2018certified,NIPS:RaghunathanSL18,CORR:abs-1805-10265}. In contrast to heuristic defenses, certified defenses provide provable guarantees against bounded adversarial perturbations.
More specifically, given a classifier and an input sample, a certified defense comes with an upper bound on the worst-case loss against norm-based attackers: the bound provides a ``certificate of robustness'', and it guarantees that no perturbation within the allowed threshold can turn the starting input into an adversarial one, therefore ruling out all attacks which are restricted to the given threshold.
Certified defenses offer a promising direction towards ending the arms race between attackers and defenders.

To improve early certification techniques, which do not scale to large dataset such as ImageNet, \emph{PixelDP}~\cite{PixelDP} leverages differential privacy to generically increase the robustness of a based classifier, offering probabilistic certificates of robustness for several datasets, including ImageNet.
This solution trades scalability with clean-data accuracy, which drops significantly as the allowed adversarial perturbation increases.
In a similar vein, recent works prove the certified robustness of \emph{randomized smoothing}, a pre-processing technique similar to~\our{} and~\regionbased{} which adds Gaussian noise to the classifier's input (instead of uniform noise)~\cite{ICML:CohenRK19,CORR:abs-1912-09899}.

On the downside, robustness certificates only hold with respect to the original input, meaning that only test inputs can be certified~\cite{CORR:abs-1902-06705}.
It is thus unclear which guarantees a robustness certificate can offer for data that has not been seen before. 
In addition, recent work by Ghiasi~\etal~\cite{Ghiasi2020BREAKING} shows an attack, so-called \emph{shadow}, to bypass certified defenses, hinting that certified robustness does not truly capture robustness to all realistic attacks. More specifically, the shadow attack generates adversarial samples which do not respect the norm-bound imposed by the certificate, and thus are technically outside the adversarial model; however, those samples are indistinguishable from the original samples, and cause the classifier to generate a ``spoofed'' certificate of robustness, ultimately bypassing the defense. 

\section{Concluding Remarks}\label{sec:discussion and conclusion}

In this work, we investigated whether \emph{randomness} can help in increasing the robustness of DL classifiers against adversarial samples.
We thoroughly analyzed three heuristic defensive techniques that employ randomness in different ways, namely \croprescale{}~\cite{ICLR:GuoRCM18}, \regionbased{}~\cite{ACSAC:CaoG17}, and \our{} (which we propose), and compared their effectiveness against state of the art graybox and whitebox attack strategies.

Our results show that randomness can enhance robustness against graybox attacks.
This is in line with prior work investigating, among others, randomized input transformations as a defense against (blackbox and) graybox attacks~\cite{ICLR:GuoRCM18}.
In contrast to prior findings~\cite{ICLR:GuoRCM18}, our results demonstrate that randomized transformations can be effective even when applied in a purely online fashion, i.e., without the need of augmenting training with the same transformations.

None of the analyzed defenses can deter whitebox attacks, though.
In fact, designing secure solutions against fully-adaptive adversaries is an extremely challenging open problem~\cite{CORR:abs-1809-02104,CORR:abs-1810-01407,CORR:abs-1805-10204,CORR:abs-1901-10861}. 
Although recent works investigate certified defenses~\cite{NIPS:HeinA17,ICML:WongK18,raghunathan2018certified,NIPS:RaghunathanSL18,CORR:abs-1805-10265}, which provide provable guarantees against bounded adversarial perturbations, these solutions are limited by the prior knowledge of the inputs, i.e., only inputs that are included within the test set can be certified,
and were recently bypassed by out-of-the-model, yet realistic, attacks.

A close look at our experimental results reveals that, although the strongest currently known whitebox attacks can successfully generate robust adversarial samples against the three defenses,
increasing the amount of added randomness requires a higher number of iterations for the attack to succeed or, similarly, a larger distortion to achieve a certain success rate.
That is, larger randomness makes the task of generating adversarial samples ``moderately harder''.
This may be good enough for applications in which the attacker has limited time to generate an adversarial sample, or is restricted to small distortion.

An interesting future direction would be to further explore the effect of randomization on the attack's cost, in terms of lower bounds for the number of iterations, respectively, amount of perturbation needed to generate robust adversarial samples. We therefore hope that our paper motivates further research in this area.

\begin{acks}
The authors are thankful to Anish Athalye, Nicholas Carlini, and anonymous reviewers for helpful comments.
This paper has (partly) received funding from the European Union's Horizon 2020 research and innovation programme under grant agreement No 779852.
\end{acks}



\bibliographystyle{ACM-Reference-Format}
\bibliography{cited}


\end{document}